\documentclass[conference]{IEEEtran}
\IEEEoverridecommandlockouts
\usepackage{graphicx}
\usepackage[pdf]{graphviz}
\usepackage{tabularx}
\usepackage{longtable}
\usepackage{placeins}
\usepackage{acronym}
\usepackage{booktabs}
\usepackage{multirow}
\usepackage{soul}
\usepackage{mdframed}
\usepackage{framed}
\usepackage{multicol}
\usepackage{balance}
\usepackage{hyperref}
\usepackage{cite}
\usepackage{amsmath,amssymb,amsfonts}
\usepackage{algorithmic}
\usepackage{graphicx}
\usepackage{textcomp}
\usepackage{xcolor}

\graphicspath{ {./images/} }

\hypersetup{
    colorlinks,
    linkcolor={red!50!black},
    citecolor={blue!50!black},
    urlcolor={blue!80!black}
}

\acrodef{SoK}{Systematization of Knowledge}
\acrodef{HCI}{Human-Computer Interaction}
\acrodef{DCS}{Developer-Centered Security}
\acrodef{CSCW}{Computer-Supported Collaborative Work}
\acrodef{IFT}{Incident Fault Tree}
\acrodef{SS}{Saltzer \& Schroeder}
\acrodef{GTM}{Grounded Theory Method}

\newenvironment{principles}{\begin{mdframed}}{\end{mdframed}}
\newenvironment{takeaway}{\begin{mdframed}\noindent\textbf{Take Away.}~}{\end{mdframed}}
\newenvironment{violation}{\begin{mdframed}\noindent\textbf{Principles.}~}{\end{mdframed}}
\newcommand{\etal}[0]{et~al{.}}
\newcommand{\gvnewline}{\noexpand\n}

\begin{document}

\title{Better Call Saltzer \& Schroeder: A Retrospective Security Analysis of SolarWinds \& Log4j}

\author{\IEEEauthorblockN{Partha Das Chowdhury, Mohammad Tahaei and Awais Rashid} \IEEEauthorblockA{Bristol Cyber Security Group\\
University of Bristol, UK\\
\{partha.daschowdhury, awais.rashid\}@bristol.ac.uk}}


\maketitle

\begin{abstract}
Saltzer \& Schroeder's principles aim to bring security to the design of computer systems. We investigate SolarWinds Orion update and Log4j to unpack the intersections where observance of these principles could have mitigated the embedded vulnerabilities. The common principles that were not observed  include \emph{fail safe defaults}, \emph{economy of mechanism}, \emph{complete mediation} and  \emph{least privilege}. Then we explore the literature on secure software development interventions for developers to identify usable analysis tools and frameworks that can contribute towards improved observance of these principles. We focus on a system wide view of access of codes, checking access paths and aiding application developers with safe libraries along with an appropriate security task list for functionalities.   
\end{abstract}


\section{Introduction}
The foundations of protecting computer systems have been developed since the 70s with secure design principles such as Saltzer \& Schroeder principles~\cite{saltzer1975protection}. Subsequently these principles have underpinned secure software development guidance~\cite{ross2016systems,martinintroduction}. Nevertheless, the instances of insecurity with negative consequences are increasing with the pervasive adoption of software systems in everything. 

We are cognizant that applying these principles requires expertise and assistance. We explore the extent to which these principles were being observed -- in the context of two major instances of vulnerable software namely, SolarWinds Orion update and Log4j. The security failure in case of both of these were significant and affected many customers/users across the world. Learning from the incident investigations with respect to the principles is our means to the end of identifying processes and mechanisms for the future. They can be embedded as timely and critical interventions that can aid application developers for improved observance the principles.  

 Our research questions (RQs) were:

\begin{mdframed}
          \noindent
    \textbf{RQ1.}
          What Saltzer \& Schroeder principles were not adequately observed in case of SolarWinds Orion update and Log4j?\\
         \noindent
    \textbf{RQ2.}
          How to aid developers to observe the principles?\\
          \noindent
\end{mdframed}

To answer our RQs, we systematically analyzed the vulnerability reports on SolarWinds Orion update and Log4j using \ac{IFT}~\cite{rashidift}, to construct the attacks in terms of events and mitigation mechanisms that could have prevented them. These mitigation mechanisms lead to the pertinent principles that were not observed. \emph{Fail safe defaults}, \emph{economy of mechanism}, \emph{complete mediation} and  \emph{least privilege} emerged as the common principles that were not observed between SolarWinds Orion Update and Log4j. We, then explored the \ac{DCS} literature to identify appropriate usable tools, development practices and frameworks that can aid towards better observance of the principles. Our work complements the studies which have looked at observance of Saltzer \& Schroeder principles among developers~\cite{jaeger2021toward, chenprompts, neumann2018fundamental} and arguments for new mechanisms in the context of \emph{fail-safe defaults} in cyber-physical systems~\cite{ massacci2019deny}. \emph{Fail safe defaults} can be aided by library developers without interfering with the paradigm of the application developers. Code reviews complemented by execution of signed code can aid towards observing \emph{economy of mechanism}, while a data driven examination of code access/execution would help observe the principle of \emph{complete mediation}. The principle of \emph{least privilege} can be observed using a functionality-security matrix --- a list of positive constructive security tasks pertaining to functionalities to be implemented.

\section{Method}
\label{method}
We use the method introduced in~\cite{rashidift} to identify \emph{unknown known} security requirements for safety-critical systems. The authors developed a \emph{novel} synthesis of grounded theory and \ac{IFT}~\cite{johnson2003handbook} and focussed on a wide class of security threats known as \emph{data exfiltration}. Their method elicits incidents across system boundaries that facilitate the perpetration of an attack and potential mitigation mechanisms that could have prevented one or more incidents in the chain. 

\paragraph{Method Justification}
\ac{GTM}~\cite{glaser1967discovery} and \ac{IFT} are two important constituents of the method we adopt in this work. \ac{GTM} iteratively reduces available data into self contained non-reducible elements. The process of data collection and breaking them is both simultaneous and independent---with the emergence of new data they go through the same process of iterations to break them up. \ac{IFT} is a modeling tool to document and analyze incidents according to their causal relationship as well as the combination (if any) of events that results in further consequent events. \emph{Events} and \emph{gates} constitute two important constructs of \ac{IFT}---the former describe an incident or a condition that led to other incidents, while the latter operates on events describing their roles in the propagation/mitigation of any resultant event. The gates are powerful as they can systematically elicit conditions that could have prevented events leading to dangerous consequences.   

The method we adopt has virtues over only qualitatively analyzing the data. The distinction between rigorous data analysis and constructing them systematically using \ac{IFT} yet their complementing relationship helps to make the model informationally rich. The holistic evaluation is essential for our purposes, given that the failures we investigated involved multiple entities similar to a supply chain scenario, while the mitigation mechanisms point towards the relevant Saltzer \& Schroeder principles. Being our focal group, developers are particularly relevant for adopting the principles in their development practices and methods. 

\paragraph{Constructs}
The constructs associated with \ac{IFT}s are as follows (Figure~\ref{SolarWinds}):
\begin{itemize}
    \item \emph{Basic event} - The modelling starts with \emph{basic events} that by themselves or together with other \emph{basic} events form the foundation of the incident. These events are non-reducible into other sub-events or other causal events.
    \item \emph{Intermediate event} - These events can be explained in terms of other events and/or causes. 
    \item \emph{Undeveloped events} - These events fall in the domain of conjectures that can be based on other events, but we have refrained from doing so in this paper. 
    \item \emph{AND gate} and {OR gate} - When multiple events come together with leading to an outcome, they are joined using the \emph{AND gate}, and if there is a disjunction, then it is denoted with a \emph{OR gate}. 
    \item \emph{Inhibit gate} - These are mitigation mechanisms that could have prevented a malicious actor from exploiting vulnerable systems. 
\end{itemize}

\paragraph{Analysis}
We qualitatively modeled the \ac{IFT}s for SolarWinds Orion update and Log4j vulnerability based on data---this data comprised SolarWinds Orion update malware analysis reports by the Cybersecurity and Infrastructure Security Agency (CISA)~\cite{cisag} of the United States government. For Log4j, we have gathered data from CISA, the Swiss Government Computer Emergency Team~\cite{swissg}, and the National Cyber Security Centrum (NCSC-NL), Netherlands~\cite{nlg}. 

We searched the malware analysis and documentation to identify the \emph{basic events} and subsequent events, which was done recursively unless the events were non-reducible and/or a causal factor was not identifiable. The events that emerged from the \ac{IFT} for each of them helped us understand the process, practices, or lack of them that led to those vulnerabilities. 

The evolution of \ac{IFT}s gave space to deliberate and identify remedial actions that could have prevented a particular security breach. These were mentioned in the \emph{inhibit gates} as the conditional event(s). These remedial actions could be physical protection mechanisms and processes. We cited relevant literature while identifying the remedial actions. While we identified the mitigation mechanism, we identified the relevant Saltzer \& Schroeder principle---a web application firewall is related to the principle of \emph{complete mediation}, which was done by comparing the descriptions of the principle in the text~\cite{saltzer1975protection} and the mitigation mechanism by the individual authors of this paper. We then identified common principles that were not observed between the SolarWinds Orion update and Log4j.   

\paragraph{Relevant Principles}
The  principles we refer to in this paper can be summarized as:
\begin{principles}
\begin{itemize}
    \item \emph{Economy of mechanism} suggests inspection of every line of code along with identification of unusual access paths, if any.
    \item \emph{Complete mediation} requires that every access should be mediated.
    \item \emph{Fail safe defaults} mandates that every access decision should be on permission.
    \item \emph{Least privilege} mandates that each program, process, or the user should only have the privileges they need to accomplish their stated task. 
\end{itemize}
\end{principles}

\section{Analysis}
\label{analysis}

\subsection{SolarWinds Orion Update}

\begin{figure*}
\label{fig:frame3}
\centering
  \includegraphics[width=1\textwidth]{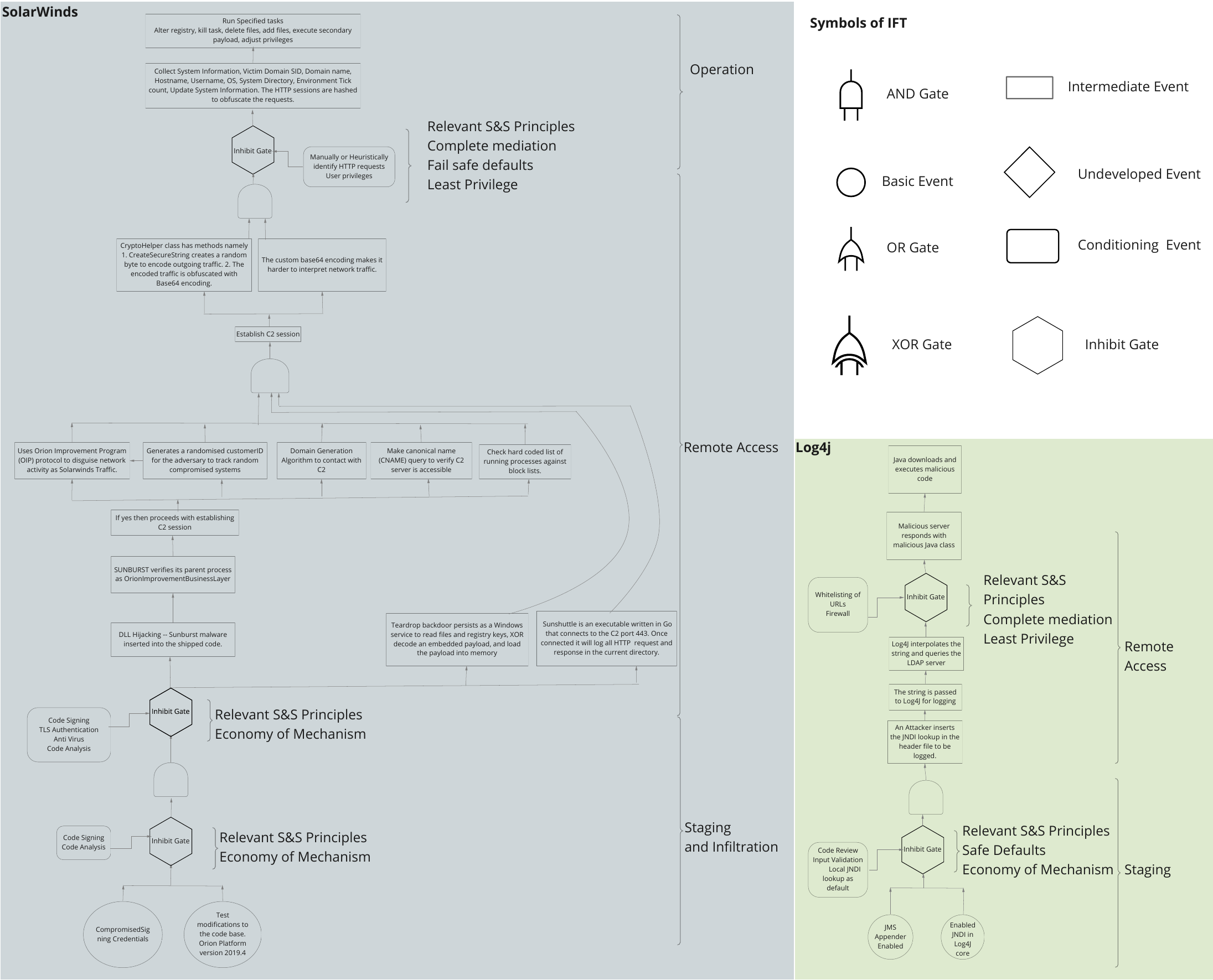}
\caption{Simplified \ac{IFT} for SolarWinds Orion Update and Log4j}
\label{SolarWinds}
\end{figure*}


CISA reported that the SolarWinds Orion update injected an APT into several critical governments and private sector enterprise systems~\cite{cisaed}. A joint statement by several federal agencies of the U.S. government~\cite{cisajs} classified this as a serious threat. SolarWinds issued a communication confirming the attack on its Orion platform, commenting on the persistence of the malicious code to sustain within the environment of the target host without being detected~\cite{swsa}. A simplified \ac{IFT} along with the possible  principles that were not observed at various stages is drawn out in Figure~\ref{SolarWinds}. In the following, we describe the violation of the principles in conjunction with events and their associated inhibit gates.

\subsubsection{Staging \& Infiltration}
\paragraph{Staging} The staging was possible due to the ability to ship malicious code signed with legitimate keys. SolarWinds, since then, has replaced their signing credentials. The adversary used this ability to sign malicious code to inject benign code into the SolarWinds Orion repository. In out IFT diagram~\ref{SolarWinds}, we classify these as \emph{basic events} that facilitated the subsequent events. 

The signing process excluded code review, verification, and bug finding. There are programs to check signed code for vulnerabilities, such as STONE~\cite{stonesoup}, which could have been the second level of defense. The principle of \emph{economy of mechanism} was not adequately observed in this instance, and the signed code was neither checked for accuracy nor access paths. 

\paragraph{Infiltration}
The ability to sign and ship malicious codes coupled with decoys facilitated the infiltration of malicious payloads---Sunburst along with Teardrop and Sunshuttle. Teardrop backdoor persists as a Windows service to read files and registry keys, XOR decodes an embedded payload and loads the payload into memory~\cite{cisasunb}. Sunshuttle is an executable written in Go that connects to the C2 port 443. Once connected, it will log all HTTP requests and responses in the current directory~\cite{cisasuns}.

We do not have data to say that the client machines failed to verify the source of the updates. Security mechanisms like code review and verification of updates failed, resulting in lowering the existing security level~\cite{schneiersolarwinds}. Such failures result drive from inadequate implementation of the \emph{economy of mechanism} principle. We refer to the STONE program again in this context for the client-side to verify the signed code. 

\begin{violation}
During the staging and infiltration stages, adherence to \emph{economy of mechanism} by the developers could have helped in detecting unwanted access paths and programming constructs that are not visible during normal use. Code review could have perhaps detected the \texttt{CryptoHelper} class and methods (1) \texttt{CreateSecureString} that generates random bytes to encode outgoing traffic, and (2) which was further obfuscated using \texttt{Base64}. 
\end{violation}

\subsubsection{Remote Access}
Once inside a target, Sunburst remains dormant before connecting to its C2 server. Before doing so, it verifies that its parent process is \texttt{OrionImprovementBusinessLayer}. The network traffic is masqueraded as Orion improvement program protocol along with intermediate events like checking for accessibility with \textit{api.solarwinds.com} to check for network connectivity. The backdoor detects any security processes and, if any are found, aborts any C2 session. Once safe, the backdoor would use a domain generation algorithm (DGA) to connect with C2 with a randomized customer ID for the adversary to track compromised systems. For example, a process such as  \texttt{GetNetworkAdapterConfiguration()} could gather network information and communicate it to the remote malicious C2.

There are inadequate implementations of multiple Saltzer \& Schroeder principles that led to an unhindered connection to remote C2. The source of requests was not verified even if they pretended to be legitimate SolarWinds traffic. The trust in requests masquerading as SolarWinds traffic was pervasive. If we combine the principle of \emph{fail safe defaults}, then the default for all outgoing traffic requests would be lack of access. 

\begin{violation}
The access of any process spawned by \texttt{OrionImprovementBusinessLayer} to \texttt{avsvmcloud.com} was neither flagged nor denied by default. This is against the principles of \emph{complete mediation} and \emph{fail safe defaults}. Sunburst verifies that the C2 server is accessible before beginning its command and control session. 
\end{violation}

\subsubsection{Operation}
Once the connection with the remote C2 is established along with a unique ID of the victim for the adversary to refer to, Sunburst collects system information, such as victim domain SID, domain name, hostname, user name, OS, and system directory. The collected information enables the remote adversary to run specified tasks, such as altering the registry, killing, deleting files, adding files, executing secondary payload, and adjusting privileges. 

SolarWinds monitors and manages on-premise and hosted infrastructures. The client-side configuration of SolarWinds Orion violated multiple principles. SolarWinds Orion was configured with pervasive privileges across the users and services, which violated the principle of \emph{least privilege}. So the breach affected many users and services and added those users to local administrators, which enabled them to install and run applications. For example, altering registry keys and disabling security processes potentially led to broader compromise. 

A complementary mechanism of \emph{reductive security} was highlighted by the authors in~\cite{schneiersolarwinds} while commenting on the SolarWinds breach, which essentially recommends stripping systems of unnecessary services, features, and libraries. This APT allowed ex-filtration of emails, log keystrokes, escalation of privileges, and deploy malicious programs. The benefits outweigh the costs. The costs involved stealing credentials at the update server, and subsequently, the malicious \texttt{dll} and loaders maintained the persistence. Lack of adherence to multiple principles significantly lowered the \emph{work factor} as well.

\begin{violation}
The methods \texttt{ExecuteEngine()},\texttt{KillTask()}, \texttt{Reboot()}, \texttt{SetProcessPrivilege()} could execute on victim taking advantage of pervasive privileges contrary to the principles of \emph{least priviledge} and \emph{separation of privilege}. Their effect was wide as the configurations violated the principle of \emph{least common mechanism}. A cumulative effect of the inadequate implementations of principles decreased the \emph{work factor} for an attacker. 
\end{violation}

\subsection{Log4j Vulnerability}

A zero-day exploit targeting the popular Java library Log4j was classified with a threat score of 10 towards the end of 2021~\cite{mitrelog4j}. Log4j is a popular Java library used for logging services. CISA, NCSC-NL, and other government agencies worldwide responded with advisories to contain the fallout of this exploit. A simplified \ac{IFT} along with the possible inadequate implementations of the Saltzer \& Schroeder principles at various stages is drawn out in Figure~\ref{SolarWinds}.

\subsubsection{Staging}
The Java Naming and Directory Interface (JNDI) provides naming and directory services to access any directory service in a common way~\cite{oraclejndi}. The property to set for a JNDI lookup is as \texttt{log4j2.enableJndiLookup=true}. When enabled the key will be prefixed with \texttt{java:comp/env/}, or with a \texttt{:} no prefix is required. The syntax for accessing a JNDI endpoint is as \texttt{jndi:logging/context-name}. Log4j publishes events to output known as appender. \texttt{JMSAppender} publishes serialized events to \texttt{JMS} topics and can lead to code execution. We classify the enabled JNDI and JMSAppender as base events in our \ac{IFT} in Figure~\ref{SolarWinds}.
 
There are key elements surrounding development and configuration practices that enable the staging of this vulnerability. The vulnerabilities for LDAP have been widely reported~\cite{mitreldap}. They include remote code execution, which is reported in the context of the exploitation of Log4j vulnerabilities in late 2021. 

Srinivasa~\etal{} conducted a honeypot study of the LDAP and Log4j vulnerabilities; they reported pivoting attacks exploiting the Log4j vulnerability~\cite{srinivasa2022deceptive}. We argue that the configurations violated the principle of \emph{safe defaults}. JNDI used JMSAppender, which is vulnerable to arbitrary code execution~\cite{jms}. A related point on versioning practices, particularly with Log4j, has been highlighted in~\cite{backes2016}; log4j lacked in their specification of version changes. 

\begin{violation}
\emph{Safe defaults} for log4j by the library developers would have prevented \texttt{JMSAppender} from executing arbitrary code after an LDAP query. Developers using the logging library perhaps could have been careful to check if a logging tool needs to execute \emph{arbitrary} code. \emph{Economy of mechanism} principle mandates a review of lines of code which was not adhered to by the developers. That could have led to input validation checks.
\end{violation}

\subsubsection{Remote Access}
The absence of input validation checks and code reviews enables an adversary to insert a malicious server as part of the JNDI lookup. The string is then passed to Log4j, which queries the LDAP server. The malicious LDAP server responds with code which Java then executes. We draw upon some of the reasons from the NCSC-NL's published list of affected systems~\cite{nlreasons}. One of the affected systems' vendors did not implement a firewall between the system and the network~\cite{abottfirewall}. A scrutiny of the list reveals that a web application firewall is one of the most advised mitigations suggested by the respective software vendors, along with updates to the library. 
 
The configuration at the client-side violated the principle of \emph{complete mediation}. The principle states that the source of every request must be checked for authority. Though the source being a logging tool should have been prevented, the remote address was not also checked for authority. Mechanisms like whitelisting could have been another layer of defense. 

The principle of \emph{least privilege} recommends that every program should operate with minimum privileges required to perform its task. The JNDI lookup enabled remote servers without input validation checks and gave the logging tool unhindered access. A vulnerable server downloads obfuscated string (an example of a firewall JNDI regex guide is as \texttt{\${\${::-j}nd\${upper:ı}:rm\${upper:ı}://127.0.0.1:\\1389}}; then the application downloads the malicious code from the malicious server. The NCSC-NL's list reports around 400 malware associated with Log4j downloaded through remote access by the logging tool~\cite{nlblob}. Threatfox database reports 186 instances of compromise as of 09/03/2022~\cite{threatfoxlog4j}.

\begin{violation}
The \texttt{jndi:ldap:maliciousserver.xxx} is a result of lack of \emph{complete mediation}. Once the LDAP points towards the malicious code's directory, Java deserializes and executes the code. Depending on the application using the library, the malicious code gained privileges in violation of the principle of \emph{least privilege}.    
\end{violation}

\section{ Implementations of Saltzer \& Schroeder principles \& Secure Software Development Practices}\label{discussion}
The \ac{IFT} evolution, along with mitigation mechanisms, reveals a list of common principles that were violated in the case of both SolarWinds Orion update and Log4j. This list is not exhaustive and is based on the data that we could gather. The  instances of inadequate implementations of the principles are summarized in table~\ref{tab:violations}.

\begin{table}
\centering
\caption{ Common inadequate implementations of the principles for SolarWinds and Log4j}
\begin{tabular}{ll}
\toprule
\textbf{SolarWinds Orion Update} & \textbf{Log4j}       \\ \midrule
Fail safe defaults               & Fail safe defaults   \\
Economy of mechanism             & Economy of mechanism \\
Complete mediation               & Complete mediation   \\
Least privilege                  & Least privilege      \\ \bottomrule
\end{tabular}
\label{tab:violations}
\end{table}

In this section, we draw from the literature on \ac{DCS} to highlight critical mechanisms that could have helped to implement the principles and, in turn, addressed the vulnerabilities and their persistence. 

\paragraph{Fail Safe Defaults}
There are notable recommendations like Green \& Smith that suggest libraries should come with safe defaults~\cite{smithgreen2016}. Das Chowdhury~\etal{} reviewed 72 peer-reviewed \ac{DCS} publications and report the \emph{challenges} faced by application developers and the \emph{behavior} they adopt to address the \emph{challenges}~\cite{partha_das_chowdhury_developers_2021}. \emph{Miscommunication} is a major challenge that developers face which arises due to a lack of documentation on how to use a library. The \emph{fail safe defaults}  principle could have been adhered to by developers using Log4j---Log4j2.16.0 (Java 8) and 2.12.2 (Java 7) disables JNDI functionality by default~\cite{nistlog4j}. We refer to the Backes~\etal{} study above, which reports the complex versioning practices of the Log4j library~\cite{backes2016}. 

From a safety perspective, if a library needs to execute arbitrary code as part of its features is a design issue, Bloch underscores the importance of problematic misuses, which perhaps can be prevented by doing away with features that are not integral to a library~\cite{bloch2006}. 

The pertinent question is how to aid developers in adhering to the principle of \emph{fail safe defaults}. Tiefenau~\etal{} conducted a randomized control trial to evaluate \texttt{Let’s Encrypt} and \texttt{Certbot}. They also situate automation and safe defaults as two key ingredients of usability while acknowledging the difficulty of automation in many cases (e.g., firewall configuration~\cite{tiefenausafe}). Their study reveals the benefits of automation with the configuration of HTTPS. They also recommend that the usable security community can learn from the \texttt{Certbot} approach for use cases like Email encryption. Gorski~\etal{} investigates the extent to which developers can be aided with implementing safe defaults. Their study with developers shows that embedding content security policy as default has positive consequences. 

\begin{takeaway}
Libraries should automate and set safe defaults; however, developers should be able to handle the edge cases. Defaults should not interfere with the developers' paradigm. For example, a security configured IDE might set a standard---Django is configured as part of the IDE for password storage. 
\end{takeaway}

\paragraph{Economy of Mechanism}
The SolarWinds Orion update had no mechanism to test the signed code before it was shipped. The unwanted access paths the signed malicious code could have taken might have been revealed through proper review before shipping the update. The processes to gather, alter system information and execute arbitrary code from malicious servers slipped without being detected. Software engineered with Log4j, if subjected to code review, could have revealed the lack of input validation checks in their implementation and usage of a library~\cite{weir2021}. Kabinna~\etal{} reviewed logging migration efforts by developers; their studies report that 70\% of applications encounter two bugs \emph{on an average} after migration and performance is \emph{rarely} improved post migration~\cite{kabinna2016logging}. Zhenhao~\etal{} report logging code smells resulting from using re-used code---often used if duplicate codes result in inaccurate logging behavior~\cite{li2019dlfinder}. 

Some studies looked at developers' use of static code analysis~\cite{johnson2013don} and code review~\cite{assal2018}. A structured code review contributes positively to developer satisfaction and application robustness yet without costing disproportionately with respect to time~\cite{codereviewgoogle}. We have discussed the STONE~\cite{stonesoup} approach in the context of \ac{IFT} for SolarWinds in section \ref{analysis}. There are standardization efforts in software procurement in the form of a software bill of materials~\cite{whitehousesbom}. This requires a complete enumeration of software and library components used in the build. 

\begin{takeaway}
Source code reviews along with static code analysis tools can help embed \emph{economy of mechanism}. They should be complemented with the execution of signed code as a standard and enumeration of components in the build process. 
\end{takeaway}

\paragraph{Complete Mediation}
While \emph{fail safe defaults} and \emph{economy of mechanism} can expose insecure access paths and input validation vulnerabilities, the failure to adhere to the principle of \emph{complete mediation} would exacerbate the insecurities. While there can be a divergence of views on the need for a logging tool to connect a remote LDAP server and have the ability to execute arbitrary code, there can be little justification for not validating the source and destination of such remote access. In the absence of safe defaults of the library and a lack of input validation check of the application code, a validation of the access request could have prevented a malicious remote entity from exfiltrating information. 

Notable \ac{DCS} studies report that security is not a priority for developers~\cite{acar2016}, so interventions discussed in~\cite{weir2021} could have led to better configuration management of the applications. These interventions range from active pen-testing of code to security awareness. Adelyar~\etal{} identify that continuous changes to software hinder a system-wide view of the software~\cite{adelyar2016towards}---this has consequences in implementing the principle of complete mediation. Taly~\etal{} builds upon the ECMA Standards committee Javascript tool and builds a tool to detect and prevent compromise of an API. Research in the future can consider building upon such approaches to have a system-wide view of programs/processes even when there are continuous changes to software~\cite{taly2011}. How to aid application developers in having a system-wide view? Erlingsson proposes a data-driven model for assurance---an essential consideration of this model is to consider the historical access/execution of a program/process. This is pertinent in the context of monitoring the continuous changes to software and their consequences~\cite{erlingssondata}. 

\begin{takeaway}
A data-driven assurance process for constituent programs/processes aggregated from various sources would reveal anomalies against their historical behavior. This should involve all execution traces.  
\end{takeaway}

\paragraph{Least Privilege}
In the case of the SolarWinds network, administrators configured SolarWinds Orion with pervasive privilege. The economic reasons behind the development practices of SolarWinds have been discussed in~\cite{schneiersolarwinds}; it is also pertinent to draw from the \ac{DCS} literature on what leads to such  inadequate implementations of Saltzer \& Schroeder principles. Assal~\etal{} interviewed developers employed in the industry to understand their security practices. The evaluation criterion is comprised of a list of security best practices which includes the principle of \emph{least privilege}. They report the importance of resourcing and division of labor in implementing the security best practices~\cite{assal2018}. If we draw from the programming practices of developers~\cite{assal2018} and extend them to the applications implementing/extending the Log4j library, this reflects upon the library developers as well. Notable \ac{DCS} studies highlight the prevalence of functionality over security~\cite{tahaei2019}---an obvious question is on the functionality of a logging function with the ability to execute arbitrary code. If that is functionality, was it tested adequately for security~\cite{linden2020}. 

Application developers need to ascertain the privileges granted to their programs/processes, which are tied to secure programming interventions like security awareness, training, penetration testing, and organizational factors such as resourcing~\cite{jose2016,weir2016}. In the context of \emph{least privilege} of programs/processes, a helpful way forward can be through \emph{prompts}. Chen~\etal{} build upon reflection as an effective way of learning; their prompts draw from the Saltzer \& Schroeder principles and parses codes for potential deviations from them. The code review then prompts the developer for any observed deviation---the prompts lead to a reflection by the developer~\cite{chenprompts}. Buyens~\etal{} identify the difficulty of implementing principles like \emph{least privilege} by developers and suggests architectural transformations---continuous splitting of components and pruning associated permissions~\cite{buyens2009}. Hammad~\etal{} developed DelDroid to detect \emph{least privilege} deviations in Android applications---they employ static analysis tools to detect privileges and take corrective actions to assign the correct privileges. Their study reports the advantages of DelDroid over hundreds of Android applications~\cite{HAMMAD201983}.
\begin{takeaway}
A functionality-security matrix will define the privileges that each program/process would need to accomplish its task. This, when complemented with prompts and static code analysis tools, can further evolve and calibrate the privileges. 
\end{takeaway}

\section{Conclusion}\label{conclusion}
While the fundamentals of protection mechanisms are there yet developers come with diverse security understanding. Conventional recommendations point towards designing software with security in mind~\cite{mcgraw1999software}. We build on the recommendation and explore the extent to which security can be weaved into development practices with explicit focus on better observance of the principles.   Our work in a way re-arranges the developer support eco-system keeping the Saltzer \& Schroeder principles at the heart of such support.  

Functionalities remain the focal point for application developers -- nevertheless they have security consequences. Library developers are ideally placed to configure safe modes for their libraries, this default configuration would take away the cognitive load from application developers in observing the principle of \emph{safe defaults}. Code reviews along with execution of signed code for adherence to \emph{economy of mechanism} would aid developers. There are usable frameworks which we discuss in section \ref{discussion}. A major contributor behind the vulnerabilities we study in this paper is unauthorized access behavior. Examining access paths of programs against their historical behavior would elicit anomalies and a data driven assurance would be appropriate and useful to observe \emph{complete mediation}. A usable functionality-security matrix has direct implication towards better observance of \emph{least privilege}. Future work can build upon the positional shift we propose in this paper for comprehensive interventions for developers.

\balance

\bibliographystyle{IEEEtran}
\bibliography{secdev}

\end{document}